# The impact of students' epistemological framing on a task requiring representational consistency

Alexandru Maries[1], Shih-Yin Lin[2], and Chandralekha Singh[3]

[1]*University of Cincinnati, Department of Physics, Cincinnati, OH 45221*
[2]*National Changhua University of Education, Department of Physics, Changhua 500, Taiwan*
[3]*University of Pittsburgh, Department of Physics and Astronomy, Pittsburgh, PA 15260*

The ability to flexibly transform between different representations (e.g., from mathematical to graphical representations) of the same concept is a hallmark of expertise. Prior research suggests that many introductory physics students show lack of representational consistency, e.g., they may construct two representations of the same concept in the same situation that are inconsistent with one another. In this case study, we asked students to construct two representations for the electric field for a situation involving Gauss's law with spherical symmetry (charged conducting sphere surrounded by charged conducting spherical shell). Prior research also suggests that this type of problem results in many students constructing representations that are not consistent with one another. Here we present findings from individual interviews with three students about this problem which suggest that students' lack of representational consistency may partly be attributed to the type of knowledge that the graphical and mathematical representations trigger. In the epistemic games framework terminology, the two representations students are asked to construct (mathematical vs. graphical) in the problem may lead them to play two different epistemic games. We discuss how students' epistemological framing may contribute to their lack of representational consistency.

## I. INTRODUCTION

Developing a solid grasp of a concept requires one to be able to express and manipulate the concept in a variety of representations [1]. As Meltzer puts it, a range of diverse representations is required to "span" the conceptual space associated with an idea [2]. It is therefore not surprising that many researchers have developed instructional strategies that place explicit emphasis on multiple representations [3-5] and have emphasized the importance of students becoming facile in translating between different representations of the same concept [6-8]. Much research in physics education has shown that students have difficulty in transforming from one representation to another consistently [9,10].

In this case study, we explore some possible reasons for introductory physics students' difficulties in expressing the electric field in a mathematical and a graphical representation consistently in the context of a Gauss's law problem. In particular, introductory students in a second semester calculus-based physics course were asked to find an expression for the electric field in a situation with a spherically symmetric charge distribution in four different regions and then plot the field on the coordinate axes provided (problem statement is provided in the methodology section). In other words, they were asked to express the electric field using two representations: mathematical and graphical. This problem was investigated previously in a large calculus-based introductory physics class and quantitative results were reported [11]. We found that many students constructed mathematical and graphical representations for the electric field in different regions which were inconsistent. Here, we focus on the insights obtained from qualitative interview data with three students which shed light on some possible reasons for the lack of consistency observed in the quantitative data.

## II. METHODOLOGY

The problem used in this study is the following:

A solid conductor of radius $a$ is inside a solid conducting spherical shell of inner radius $b$ and outer radius $c$. The net charge on the solid conductor is $+Q$ and the net charge on the concentric spherical shell is $-Q$ (see Fig. 1).
(I) Write an expression for the electric field in each region.
  (i)   $r < a$
  (ii)  $a < r < b$
  (iii) $b < r < c$
  (iv)  $r > c$

(II) On the figure below (see Fig. 2), plot $E(r)$ (which is the magnitude of the electric field at a distance $r$ from the center of the sphere) in all regions for the problem in part (I).

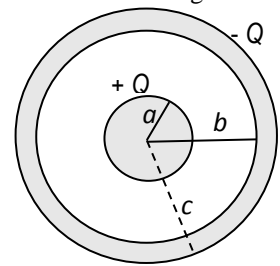

**FIG 1.** Problem diagram

In order to identify possible reasons for the inconsistency observed in the quantitative data [11], we conducted individual think-aloud [12] interviews with introductory students who had completed the study of electrostatics and had been tested on it via a midterm exam. In particular, we were interested in the paths that students may follow which lead them to obtain inconsistent answers. During the 30-60 min interviews (which were audio recorded), students were asked to solve the problem while thinking out loud. They were not interrupted during the interview except when they became quiet for a long time, in which case they were reminded to continue talking. At the time of the interviews, all of the students were enrolled in a second semester calculus-based introductory physics course and had passed the first semester calculus-based introductory physics course which they had taken at the same institution one or two semesters prior to the interviews. All the courses were using primarily lecture-based instruction.

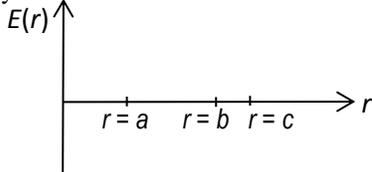

**FIG 2.** Coordinate axes provided for plotting the electric field magnitude vs. distance from the center of the sphere.

For experts, the task of finding an expression for the electric field and the task of plotting the electric field are closely related to one another and when expressing the electric field in different regions using two different representations, they would ensure that the two representations are consistent with one another in each region. Also, if they are asked to complete the two tasks separately, they are likely to use similar knowledge to complete them. On the other hand, one hypothesis for why introductory students, who are still developing expertise, may not recognize that the two representations of the electric field must be consistent in each region is that the two representational tasks may trigger students to use different pieces of knowledge to complete them. Below, we describe the findings from individual interviews with three students, which shed light on some of the reasons for why students' representations may lack consistency.

## III. RESULTS

We focus on interviews with three students: Sarah, Joe, and John. Sarah and Joe received fairly average or slightly below-average grades in the first semester course (B-, C+ respectively), while John received an A+. In the second semester course, in the first exam, Sarah and Joe (enrolled in the same section) received close to average grade, while John received well above average grade.

Interviews suggest that for some students, asking them to find an expression for the electric field in the given situation triggers them to use the mathematical form of Gauss's law or try to recall a mathematical expression for the electric field. On the other hand, when asked to plot the electric field, the same students may think qualitatively about Gauss's law in order to determine the behavior of the electric field ($E=0$ or proportional to $1/r^2$, etc.).

Due to the common introductory student difficulty in applying the mathematical form of Gauss's law correctly [13], we found that more often than not, students' plots for the electric field in different regions in Fig. 2 (obtained from qualitative reasoning) did not agree with their mathematical expressions (obtained from recalling a mathematical expression or applying the mathematical form of Gauss's law). Adopting the epistemic games framework [14], some students appeared to be playing different epistemic games when finding an expression for the electric field and when plotting the field. For example, when finding an expression for the field, some students appeared to be playing the Mapping Meaning to Mathematics (MMM) epistemic game [14] in which they started by developing a story about the physical situation, then translated the story into mathematical entities, performed mathematical steps and finally evaluated the story. However, when plotting the electric field, the same students appeared to play the Physical Mechanism (PM) epistemic game [14] in which they attempted to construct a story based on their intuitive sense of the physical situation without explicit use of physics equations relevant for those situations. For example, in region $r<a$ Sarah started by thinking qualitatively and said:

*Sarah: "I will need to consider […] Gauss's law, and for a conductor, I don't have to consider it [the charge] […] It's only in an insulator where the charge doesn't all distribute to the surface, so the charge [in this situation] for r less than a should be zero. So the electric field should be zero."*

She appeared to play the PM epistemic game (no charge enclosed, therefore the electric field is zero). We note that Tuminaro and Redish define the knowledge base of the PM game to be consisting entirely of reasoning primitives. Here, we expand upon it by also including instances in which students reason qualitatively without referencing equations, as none of the other epistemic games they discuss appear to include such instances.

After Sarah concluded that the electric field should be zero by using the PM epistemic game, she started playing the MMM epistemic game. She incorrectly wrote down that $E = qF$ (she was trying to recall the connection between electric field and electric force, namely, $\vec{F} = q\vec{E}$) and then stated that because the charge enclosed is zero in this region (i.e., $q=0$ in the equation she wrote down), $E$ will be zero. She trusted her mathematical approach (although incorrect) more than her qualitative reasoning (which she used while plotting) for why the field is zero in that region. She explicitly commented that she was not confident that the equation she remembered, $E = qF$, was correct, but it appeared from the interview that she trusted it more than her qualitative reasoning. She only wrote down $E=0$ after she could obtain it from this mathematical approach with $q=0$. At a later point in the interview, she briefly looked back at her work in this region, paused for a few seconds, perhaps because she was unsure whether the equation she recalled ($E = qF$), was correct. She then noted, "either way, you get zero", which indicated that she was aware that she solved this part with different approaches, both of which yielded $E=0$.

In region $b<r<c$ (also within a conductor, where the electric field is zero), Sarah also started by reasoning qualitatively (she appeared to be playing the PM epistemic game [14]) and stated:

*Sarah: "In there, it should be zero because it's within a conductor."*

Further discussions with Sarah suggest that she was not sure why the electric field within a conductor should be zero and she was not able to refer to physics principles or equations that would ensure that the field is zero inside a conductor. She was just using her intuitive sense about the physical situations involving conductors to construct the story that the electric field in that region is zero. However, similar to her approach in the first region, Sarah did not write down *E*=0 for the field in that region, and instead tried to apply Gauss's law to find the electric field (she appeared to play the MMM epistemic game [14]). However, she did so incorrectly and obtained $E = -4\pi c^2 + 4\pi b^2$, after which she explicitly stated:

*Sarah: "The electric field will be equal to negative four pi c squared minus four pi b squared. And it will be equal to zero (our emphasis). I just know that."*

When plotting the field in this region, Sarah plotted a zero electric field, even though her expression was non-zero. To Sarah's credit, she *was* worried that her expression may not be consistent with what she plotted. In other words, since the two epistemic games that she played did not lead to consistent descriptions of the field in the region, she was concerned about it and stated:

*Sarah: "Hmm… that's not always gonna work out…that four pi c squared and four pi b squared will cancel out in the equation to give zero […] But I don't have anything better in my head right now."*

This interview suggests that Sarah was more inclined to trust an expression for the electric field if she found it by using a mathematical procedure (playing the MMM epistemic game). In particular, it appeared that she trusted $E = -4\pi c^2 + 4\pi b^2$ as her expression for electric field (because this is what she wrote down) instead of *E* = 0 which is what she plotted in region *b*<*r*<*c*. Also, in region *r*<*a*, she only wrote *E*=0 as her expression after she obtained this expression by using a mathematical procedure (*E* = *qF* with *q*=0). In both of these regions, when plotting the field, she did not go back to look at her expressions and plotted the behavior she was expecting from her qualitative reasoning about conductors, (i.e., from playing the PM epistemic game [14]), which, in both regions, was *E* = 0. However, in region *b*<*r*<*c,* her plot was inconsistent with her expression.

After the interview, when Sarah was asked why she wrote $E = -4\pi c^2 + 4\pi b^2$ instead of *E* = 0, despite her expression and plot being inconsistent with each other, she stated:

*Sarah: "Sometimes, I need the conceptual to pull me into the math, but when they don't line up, […] you just have to go with the math."*

Further discussions with Sarah suggest that her epistemological beliefs [15] may be hurting her problem solving performance. In particular, she believed that her intuition did not necessarily need to be reconciled with the mathematical formalism and that quantitative reasoning should be trusted more when it does not match with qualitative reasoning when finding an expression for the electric field. These beliefs may have prevented her from back tracking and performing a reasonability check to ensure that the two representations were consistent.

Sarah employed a similar approach in region *a*<*r*<*b*, in which she obtained, $E = -4\pi b^2 + 4\pi a^2$, by playing the MMM game. But when plotting the electric field, she stated:

*Sarah: "For r between distances a and b […] we dropped off with E being proportional to 1/r².*"

She then plotted a function that decreases as inverse square instead of plotting the expression she found through mathematically applying Gauss's law (a constant negative function). It is not entirely clear from the interview why Sarah thought (correctly) that the electric field between the inner sphere and outer spherical shell should be proportional to 1/$r^2$. It is possible that she was, to a certain degree, playing the PM game, because her intuition was telling her that the electric field should decrease in this manner away from the surface of the sphere). Whatever the actual reason, she plotted this *1/$r^2$* behavior instead of the mathematical expression for the field she obtained.

Thus, when the two epistemic games resulted in the same answer (region *r*<*a*), she was most confident. However, when they yielded different answers, she trusted what she obtained by playing the MMM epistemic game when writing down an expression for the field (as she did in regions *a*<*r*<*b* and *b*<*r*<*c* discussed earlier), but when plotting the field, she plotted the behavior indicated by the PM epistemic game.

Another student, Joe, also appeared to play these two epistemic games which resulted in different answers in the same region using the two approaches. He wrote down an expression he found by applying Gauss's law mathematically or recalling a memorized result (he did not explicitly show his work to arrive at his expression starting from Gauss's law), but plotted the behavior he was expecting from qualitative reasoning. For example, in region *r*<*a*, he first recalled and wrote down the formula for the electric field outside of a conducting spherical shell carrying a total charge *Q*, namely, *E*=*kQ*/$r^2$ . However, he stated that inside the sphere you cannot use this formula and you have to use volume charge density. Therefore, inside the sphere, he wrote down $E = \frac{k\rho}{r^2}$ (in this expression, *ρ* refers to volume charge density, which Joe did not define explicitly). On the other hand, when plotting the electric field, he plotted a zero electric field and stated:

*Joe: "There's no field here – it's zero from r equal zero to r equals a"*

Discussions with him did not clarify why he thought that the electric field is zero in this region. However, based on his approach in subsequent regions, in which he plotted what he expected from qualitative reasoning, he may have been doing the same for this region.

Furthermore, in region $b<r<c$, Joe found a non-zero mathematical expression, $k|Q||\rho|/r^2$. However, when he plotted the electric field in this region, he stated:

*Joe: "There's gonna be no electric field inside this region because the charges [−Q] are all on this [inner] surface."*

Thus, while plotting the field, he used correct qualitative reasoning to arrive at the conclusion that the electric field is zero (i.e., employed the PM epistemic game productively [14] to plot the field in this situation). On the other hand, when writing down an expression for the electric field, he trusted the mathematical expression (field proportional to $1/r^2$) that he found and did not modify it despite the mismatch between the plot and the expression.

Another student, John, appeared to employ similar approaches: in region $a<r<b$, he appeared to play the MMM game in order to find the expression for the field [14]. He first used an expression he recalled for the field at a distance $r$ away from a sphere of charge $Q$ ($E = kQ/r$) and included contributions both from the inner sphere and the outer spherical shell to obtain $2kQ/r^2$ for the field in that region. However, while plotting, John stated that the physical situation given in the problem was a spherical capacitor and intuitively argued (by playing the PM game) that the field should be constant:

*John: "As we get farther away from [the edge of the sphere], [...] the outer circle's [outer spherical shell] field would get stronger in a way that the [net] field would remain constant anywhere between the two points [r=a and r=b]."*

John plotted a constant, positive electric field between $r=a$ and $r=b$ (obtained from PM game) instead of plotting the function he wrote down for the field in this region (~$1/r^2$). Thus, John also plotted his "expected" behavior instead of his expression similar to Sarah and Joe.

We note that one could have interpreted the qualitative data presented here from the lens of formal vs. informal reasoning [16]. However, this would result in similar insight.

## IV. SUMMARY

In this case study, we analyze think-aloud interviews with three students to identify possible reasons for students' inconsistency between a mathematical and a graphical representation of the electric field for a situation involving spherical symmetry of charge. We found that students wrote expressions they obtained for different regions but often plotted their intuitive answer (which did not necessarily match the expressions) instead of attempting to reconcile the two different answers. For example, students sometimes were aware that the field vanished in a particular region by thinking qualitatively (i.e., playing the physical mechanism epistemic game). However, instead of writing down $E = 0$ in that region, they appeared to play the mapping meaning to mathematics epistemic game to find an expression for the field. This often led them to obtain expressions inconsistent with the ones obtained using qualitative reasoning. Also, students encountered a similar difficulty in the region which had a non-zero field for which the mathematical procedure did not result in their "expected" behavior. Being unable to reconcile the two approaches, qualitative and mathematical, students often trusted the expression found by playing the mapping meaning to mathematics epistemic game more. However, when plotting the field, instead of plotting the mathematical expression they wrote down, they plotted the "expected" behaviors obtained from playing the physical mechanism epistemic game. Thus, playing different epistemic games can at least partly account for the lack of consistency observed in the quantitative data [11]. Thus, it appears that some students were inconsistent between their expressions and plots for the field not because they did not know how various functions were supposed to be plotted, but because they did not plot those functions. Instead, they plotted other functions obtained through qualitative reasoning. Thus, the physics context in which the plotting task was encased may have had a detrimental impact on their representational consistency and caused them to play different epistemic games when writing an expression for the field and plotting it.

## ACKNOWLEDGEMENTS

We thank NSF for award Phy-1505460.